# On-Chip Emitter-Coupled Meta-Optics for Versatile Photon Sources

Sören im Sande, Yinhui Kan, Danylo Komisar, Cuo Wu, Shailesh Kumar, Sergey I. Bozhevolnyi, Fei Ding*

Centre for Nano Optics, University of Southern Denmark, Campusvej 55, Odense M DK-5230, Denmark

*Corresponding author: feid@mci.sdu.dk

**Abstract**

Controlling the spontaneous emission of nanoscale quantum emitters (QEs) is crucial for developing advanced photon sources required in many areas of modern nanophotonics, including quantum information technologies. Conventional approaches to shaping photon emission are based on using bulky configurations, while approaches recently developed in quantum metaphotonics suffer from limited capabilities in achieving desired polarization states and directionality, failing to provide on-demand photon sources tailored precisely to technological needs. Here, we propose a universal approach to designing versatile photon sources using on-chip QE-coupled meta-optics that enable *direct* transformations of QE-excited surface plasmon polaritons into spatially propagating photon streams with arbitrary polarization states, directionality, and amplitudes *via* both resonance and geometric phases supplied by meta-atoms. Leveraging this platform, we experimentally demonstrate the independent engineering of polarization states, directionality, and relative amplitudes of photon emission in single and multiple radiation channels. The developed universal approach represents a major step towards the realization of versatile quantum sources, highlighting the potential of integrated photon sources with superior properties for both classical and quantum information technologies.

**Introduction**

Direct use of solid-state quantum emitters (QEs), such as quantum dots, molecules, and defects in nanodiamonds, encounters several challenges due to poor QE emission characteristics, including low emission rates, omnidirectionality, and insufficiently defined polarization states, a circumstance that hinders their applications in photonic and quantum technologies [1]. Typically, shaping the emission of QEs requires bulky and heavy components (e.g., polarizers, wave plates, and spatial light modulators) that are not compatible with further developments in large-scale on-chip photonic quantum systems [2-4]. To circumvent these issues, QEs have been integrated into flat optics, i.e., with quasi-flat nanostructures offering more compact solutions for advanced functionalities [5-10]. For instance, nonradiative coupling QEs to nanocavities in the near field leads to higher emission rates through the Purcell effect, although encumbering full control over polarization states and directivity [11-18]. Metalenses have been employed as external functional components to manipulate the QE emission properties in the far field, extending control over directionality and polarization but often resulting in complicated alignment and enlarged thicknesses [19-22].

Recently, on-chip QE-plasmon coupled quantum metasurfaces have been proposed for emission engineering, benefiting from planar configurations, simple fabrication processes, and possibilities to control at least some degrees of freedom (DoFs) [23-29]. These hybrid metasurfaces enable the excitation of propagating surface plasmon polaritons (SPPs) that are eventually outcoupled by surrounding meta-atoms, consisting of either dielectric nanostructures [23-27] or nanoslits in metal films [28,29], resulting in circularly polarized (CP) emission [23-26,28,29], orbital angular momentum (OAM) encoded emission [24-26], as well as linear polarization generation and splitting [27]. However, currently exploited quantum metasurfaces possess limited capabilities in simultaneously achieving desired polarization states and directionality for photon emission, not to mention multichannel emissions with arbitrary DoFs, be it with using the scattering-holography [24,25,27] or direct-scattering [26,28,29] approach.

In this work, we present a universal strategy to realize versatile photon sources with arbitrarily designed polarization states and directionality by directly transforming QE-coupled outgoing radially polarized (RP) SPPs into well-collimated propagating beams with all accessible DoFs. This major development is achieved by precisely designing meta-atoms with spin-decoupled scattering phases governed by both structural resonance and geometric origins. We demonstrate highly directional and off-normal single-mode emission with pure linear and circular polarization states, as well as angularly separated multichannel emissions with orthogonal polarization states. Finally, we highlight the versatility of our approach by generating simultaneously linear and circular polarizations with a single metasurface and demonstrating the design freedom in adjusting their relative intensities.

**General concept for designing versatile photon sources with on-chip QE-coupled meta-optics**

The general design concept is based on determining the optical properties of QE-coupled meta-optics that would transform QE-excited in-plane SPP waves $\vec{E}_{\text{SPP}}|\sigma_{\text{SPP}}\rangle$ into propagating photons $\vec{E}_{\text{ph}}|\sigma_{\text{ph}}\rangle$ with freely tailored polarization state, directionality, and phase profile (Fig. 1). Specifically, the source is diverging RP SPP wave described by a Jones vector $\vec{E}_{\text{SPP}}|\sigma_{\text{SPP}}\rangle = \frac{1}{\sqrt{2}}e^{i\varphi_{\text{SPP}}}(e^{i\varphi_{l0}}|l\rangle + e^{i\varphi_{r0}}|r\rangle)$ (Fig. 1a), where $\varphi_{\text{SPP}}$ is the SPP propagation phase, $\varphi_{l0} = -\text{atan}\frac{y}{x}$ and $\varphi_{r0} = \text{atan}\frac{y}{x}$ are initial phases from SPPs with $x$ and $y$ being cartesian coordinates, and the QE being the origin, and $|l\rangle$ and $|r\rangle$ denote the left-handed and right-handed circularly polarized (LCP and RCP) states, respectively. For emitted photons with an arbitrary polarization state $|\sigma_{\text{ph}}\rangle$, directionality $\vec{\rho}$ (defined by the polar angle $\theta$ and azimuthal angle $\varphi$), and phase profile $\varphi_{\text{pha}}$, they can be described by a point $(\chi_{\text{pol}}, \psi_{\text{pol}})$ on Poincare's sphere (Fig. 1b)

$$\vec{E}_{\text{ph}}|\sigma_{\text{ph}}\rangle = e^{i\varphi_{\text{pha}}}e^{i\beta z}e^{i\varphi_{\text{dir}}}\left(\sqrt{\frac{1-\sin(2\chi_{\text{pol}})}{2}}e^{i\psi_{\text{pol}}}|l\rangle + \sqrt{\frac{1+\sin(2\chi_{\text{pol}})}{2}}e^{-i\psi_{\text{pol}}}|r\rangle\right) \quad (1)$$

where $\beta = k_0\cos\theta$ is the propagation constant along the $z$-direction, $k_0$ being the free-space wavevector, and $\varphi_{\text{dir}} = k_0\sin\theta\,(x\cos\varphi + y\sin\varphi)$ denotes the in-plane phase factor of emitted photons. To outcouple RP SPPs into target photons, we consider a general reflective metasurface constructed by a set of meta-atoms, achieving the following condition (Supplementary S1):

$$e^{-i\varphi_{\text{SPP}}}M(\theta_{\text{m}})\begin{pmatrix}e^{i\varphi_{l0}}\\ e^{i\varphi_{r0}}\end{pmatrix} = e^{i\varphi_{\text{pha}}}e^{i\varphi_{\text{dir}}}\begin{pmatrix}\sqrt{\frac{1-\sin(2\chi_{\text{pol}})}{2}}e^{i\psi_{\text{pol}}}\\ \sqrt{\frac{1+\sin(2\chi_{\text{pol}})}{2}}e^{-i\psi_{\text{pol}}}\end{pmatrix} \quad (2)$$

where $M(\theta_{\text{m}})$ denotes the optical response of the meta-atom with a rotation angle of $\theta_{\text{m}}$, each possessing four degrees of freedom to get an accurate solution.

To generate linearly polarized (LP) photons with an arbitrary oscillating direction (i.e., $\vec{E}_{\text{ph}}|\sigma_{\text{ph}}\rangle = \frac{1}{\sqrt{2}}e^{i\varphi_{\text{pha}}}e^{i\beta z}e^{i\varphi_{\text{dir}}}(e^{i\psi_{\text{pol}}}|l\rangle + e^{-i\psi_{\text{pol}}}|r\rangle$ and $0 \leq \psi_{\text{pol}} < \pi$), we can simply use Ag-SiO$_2$-Ag meta-atoms that function as half-wave plates (HWPs, Fig. 1c, and Supplementary S2) to provide spin-decoupled phases $\delta_{\text{xx}} \mp 2\theta_{\text{m}}$ for $|l\rangle$ and $|r\rangle$, with $\delta_{\text{xx}}$ being the resonance phase in the linear polarization basis. If we ignore the amplitude variation of meta-atoms and the loss of SPPs, we get the following phase-matching condition

$$\begin{cases}\varphi_{\text{LCP}} = \delta_{\text{xx}} - 2\theta_{\text{m}} = \varphi_{\text{pha}} + \varphi_{\text{dir}} + \psi_{\text{pol}} + \varphi_{l0} + \varphi_{\text{SPP}}\\ \varphi_{\text{RCP}} = \delta_{\text{xx}} + 2\theta_{\text{m}} = \varphi_{\text{pha}} + \varphi_{\text{dir}} - \psi_{\text{pol}} + \varphi_{r0} + \varphi_{\text{SPP}}\end{cases} \quad (3)$$

Therefore, by properly selecting HWP meta-atoms with specific dimensions and orientations, we can generate propagating photons with an arbitrary linear polarization state $|\sigma_{\text{ph}}\rangle = e^{i\psi_{\text{pol}}}|l\rangle + e^{-i\psi_{\text{pol}}}|r\rangle$, directionality $\vec{\rho}$, and phase profile $\varphi_{\text{pha}}$. For high-purity CP photon sources, we only need to make desired CP

photons radiate while forbidding the radiation of its orthogonal CP counterpart by solely matching the radiative phase of the targeted state. Moreover, this concept can be easily extended to multiple beams with controllable intensities by superposing all components with the weight $C_j$:

$$\begin{cases} \varphi_{\text{LCP}} = \delta_{\text{xx}} - 2\theta_{\text{m}} = \arg\left(\sum_{j=1}^{n} C_j\, e^{i(\varphi_{\text{pha\_l\_j}} + \varphi_{\text{dir\_l\_j}} + \psi_{\text{pol\_j}} + \varphi_{\text{l0}} + \varphi_{\text{SPP\_l\_j}})}\right) \\ \varphi_{\text{RCP}} = \delta_{\text{xx}} + 2\theta_{\text{m}} = \arg\left(\sum_{j=1}^{n} C_j\, e^{i(\varphi_{\text{pha\_r\_j}} + \varphi_{\text{dir\_r\_j}} - \psi_{\text{pol\_j}} + \varphi_{\text{r0}} + \varphi_{\text{SPP\_r\_j}})}\right) \end{cases} \quad (4)$$

As shown in Fig. 1b, two spatially separated beams with different polarization states, directions, and intensities ($C_1 e^{-i\vec{k}_0 \vec{\rho}_1}|a\rangle$ and $C_2 e^{-i\vec{k}_0 \vec{\rho}_2}|r\rangle$) can be accordingly achieved without segmenting or interleaving the metasurface.

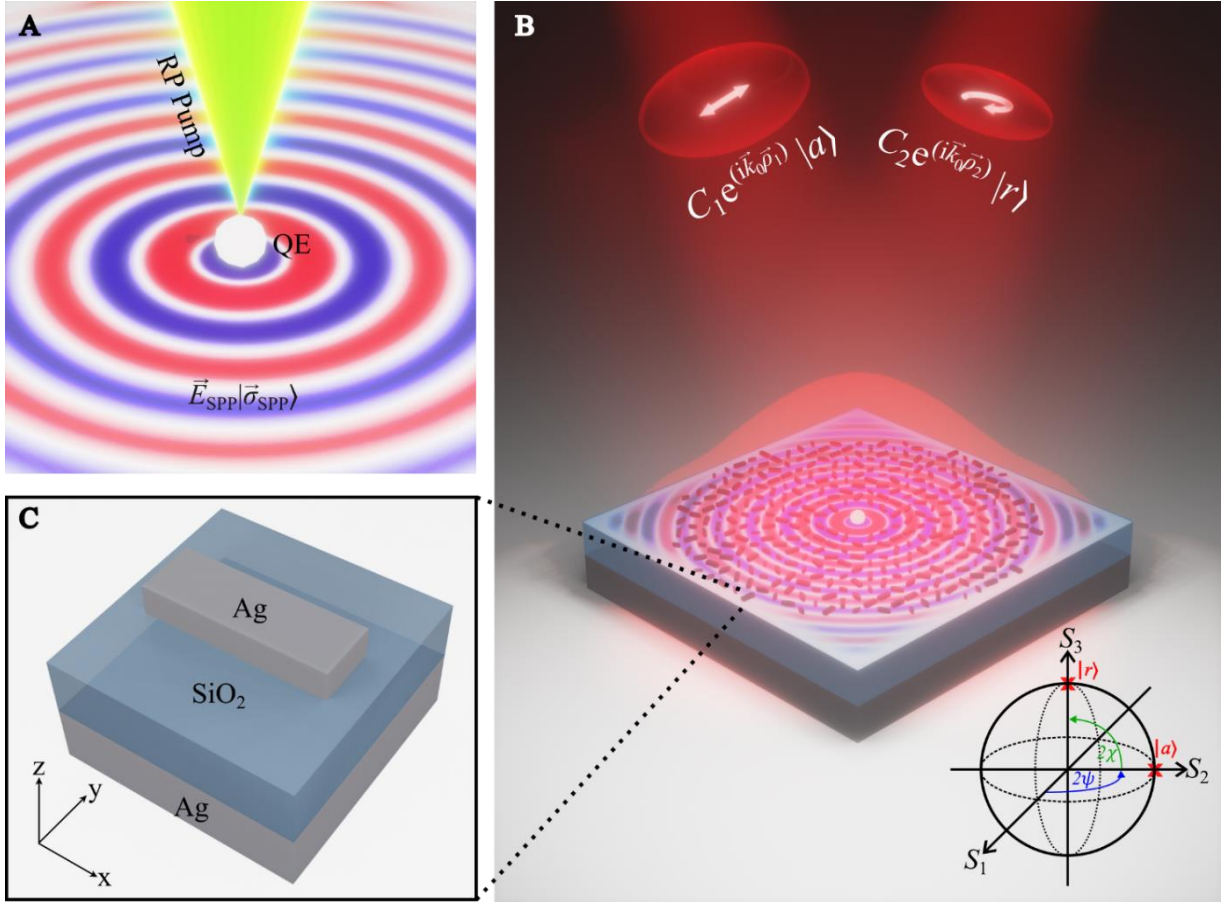

**Figure 1**: Working principle of the QE-coupled meta-optics platform for transforming QE-excited RP SPP waves $\vec{E}_{\text{SPP}}|\sigma_{\text{SPP}}\rangle$ into propagating photons with a target state of $\vec{E}_{\text{ph}}|\sigma_{\text{ph}}\rangle$. (**A**) Schematic of outgoing SPP waves excited by a QE. An RP green laser (i.e., 532 nm) is used to excite the QE, resulting in the efficient emission of diverging RP SPP waves. (**B**) Schematic of versatile photon emission. The diverging RP SPPs are scattered by the meta-atoms to produce two spatially separated beams with different polarization states, directionality, and intensities, represented as $C_1 e^{-i\vec{k}_0 \vec{\rho}_1}|a\rangle$ and $C_2 e^{-i\vec{k}_0 \vec{\rho}_2}|r\rangle$. (**C**) Schematic of a metasurface element along with the coordinate system used.

**Experimental demonstration of well-collimated, single-channel photon emission**

As a benchmark test, using silicon dioxide ($SiO_2$) coated silver (Ag) substrates, we first designed and fabricated QE-integrated metasurfaces with diameters of 10 μm and inner radii of approximately 300 nm to generate well-collimated, single-channel photon emission with desired linear polarization states and directionality (Fig. 2 and Supplementary S3 to S5). This was achieved by utilizing nanodiamonds containing NV-centers (ND-NVs) as QEs with an emission peak of 670 nm at room temperature, which were surrounded by a precisely engineered set of HWP meta-atoms exhibiting various resonance and geometric phases to satisfy the phase-matching criteria for both LCP and RCP components. Despite some fabrication deficiencies (Fig. 2a), the fabricated metasurface demonstrates excellent performance in generating collimated, single-channel $x$-polarized emission in the far field (Fig. 2b), where the emission is displaced by a polar angle of $\theta = 30°$ and an azimuthal angle of $\varphi = 135°$, as expected. The polarization-resolved far-field intensity patterns illustrate the high contrast between $|x\rangle$ and $|y\rangle$ polarization states (Fig. 2b), indicating the high purity of the emitted polarization state. This high purity is quantitatively confirmed by the measured Stokes parameter $S_1$ [defined as $S_1 = (I_x - I_y)/(I_x + I_y)$, with $I_x$ and $I_y$ being the normalized emission intensity in the $|x\rangle$ and $|y\rangle$ states, respectively], superposed with the emission pattern in Fig. 2c. Impressively, the $S_1$-parameter value remains consistently high across the entire emission spot with the averaged $\overline{S_1}$ within the full width at half maximum (FWHM) of the emission spot reaching ~0.986. To properly describe the experimental observation, the intensity-weighted Stokes parameter, defined as $S_{1w} = S_1 \times \max(I_x, I_y)$, closely follows the intensity profile over the cross-section of the spot (Fig. 2d), demonstrating the unidirectional generation of collimated $x$-polarized photons with high polarization purity. Moreover, experimental results show excellent agreement with simulation ones in terms of both the far-field pattern and Stokes parameter $S_1$ (Supplementary Fig. S2). When the designed linear polarization state is changed from $|x\rangle$ to its orthogonal state $|y\rangle$, unidirectional single-channel emission with high polarization purity (with $\overline{S_1} \approx -0.989$) is realized in the far field (Supplementary Figs. S4 and S5), showcasing the versatility of our design approach. To miniaturize the photon sources, we fabricated a smaller metasurface with a diameter of 5 μm for collimated $|x\rangle$ state generation (Supplementary Figs. S6), which is otherwise identical to the metasurface shown in Fig. 2. This smaller metasurface exhibits similarly good performance in producing unidirectional $x$-polarized emission $\overline{S_1}$ reaching approximately 0.980, thereby confirming the feasibility of miniaturization (Supplementary Fig. S6c). However, the drawback of this miniaturization is the increased spot size (Supplementary Figs. S6b and S6d). Furthermore, the inner radius between the ND-NVs and meta-atoms can be varied to optimize the emission intensity in the targeted polarization state (Supplementary Fig. S7), attributed to the feedback of the reflected SPP waves after interacting with meta-atoms. For instance, changing the inner radius from 300 nm to 1.9 μm can increase the $|x\rangle$ intensity by around 25%. This new design with a large inner radius could further alleviate the alignment requirements during fabrication. Apart from homogeneous linear polarization states, such as $|x\rangle$, $|y\rangle$, and $|a\rangle$ [$|a\rangle = \frac{1}{\sqrt{2}}(|x\rangle + |y\rangle)$, Supplementary Fig. S8], inhomogeneous linear polarization states possessing vector properties can be designed using our approach (Supplementary Fig. S9). Moreover, complex phase profiles like spiral phases can be incorporated to generate well-collimated, single-channel $|x\rangle$ emission carrying orbital angular momentum (Supplementary Fig. S10).

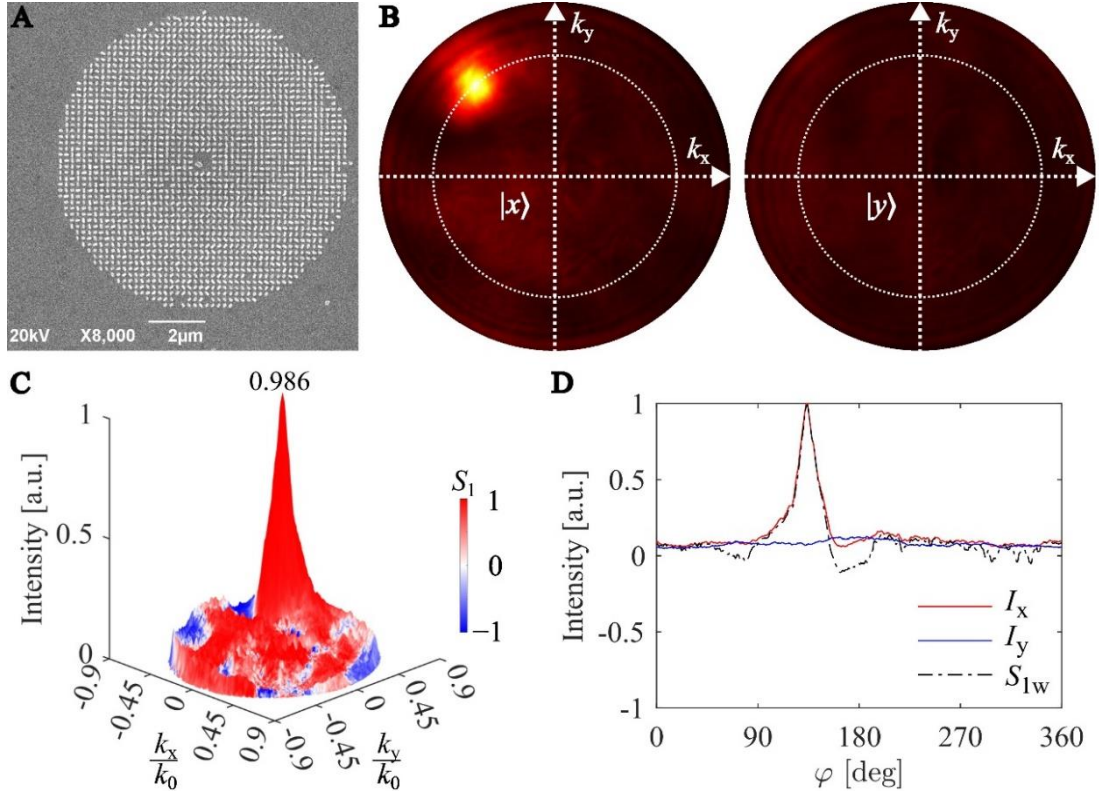

**Figure 2:** Experimental demonstration of well-collimated, single-channel emission with $|x\rangle$ state. (**A**) Scanning electron microscope (SEM) image of the fabricated metasurface with a diameter of 10 µm and an inner radius of ~300 nm. (**B**) Polarization-resolved far-field intensity distributions of orthogonally polarized emission. Dashed circles indicate the polar angle of $\theta = 30°$. (**C**) Three-dimensional (3D) representation of the superposed far-field intensity (height) and Stokes parameters $S_1$ (color). (**D**) Cross-sectional distributions of the normalized intensities $I_x$ and $I_y$ and weighted Stokes parameters $S_{1w}$ on a circular path corresponding to the polar angle of $\theta = 30°$.

To create a single emission channel with a purely circular polarization state, we only allow the desired CP photons to radiate while forbidding the radiation of their orthogonal CP counterparts by setting different radiative phases. For instance, the radiative phases of the $|l\rangle$ and $|r\rangle$ components are set as $\varphi_{\text{SPP\_l}} = k_{\text{SPP}}\sqrt{x^2 + y^2}$ and $\varphi_{\text{SPP\_r}} = 0$ (Supplementary S1), respectively, to implement a specific metasurface for unidirectional single-channel LCP emission with high fidelity, as shown in Fig. 3a and Supplementary Fig. S11. The measured far-field intensity distributions show a significant contrast between $|l\rangle$ and $|r\rangle$ states (Fig.

3b), where a pronounced $|l\rangle$ spot is collimated at the point of $(\theta, \varphi) = (30°, 270°)$ in the Fourier plane while the $|r\rangle$ component is greatly suppressed to the noise level, effectively highlighting the metasurface's capability to selectively emit the targeted circular polarization in a desired direction, superior to nanoridges suffering from coupled CP photons [23-25] and nanorod dimers with limited directionality [26]. The measured Stokes parameter $S_3$ [defined as $S_3 = (I_r - I_l)/(I_r + I_l)$, with $I_r$ and $I_l$ being the normalized emission intensity in the $|r\rangle$ and $|l\rangle$ states, respectively] reaches −0.938 once averaged over the emission spot ($\overline{S_3}$), indicating a high purity of the circular polarization state (Fig. 3c). Additionally, the high degree of circular polarization is consistently high over the cross-section of the spot, with the intensity-weighted Stokes parameter $S_{3w} = S_3 \times \max(I_r, I_l)$ remaining vertically symmetric to the intensity profile (Fig. 3d). As expected, the fabricated metasurface shows excellent agreement with the simulation results, validating the design approach (Supplementary Fig. S11). The calculated EQE for this metasurface design reaches ~0.5, demonstrating its effectiveness and efficiency in generating the desired CP emission.

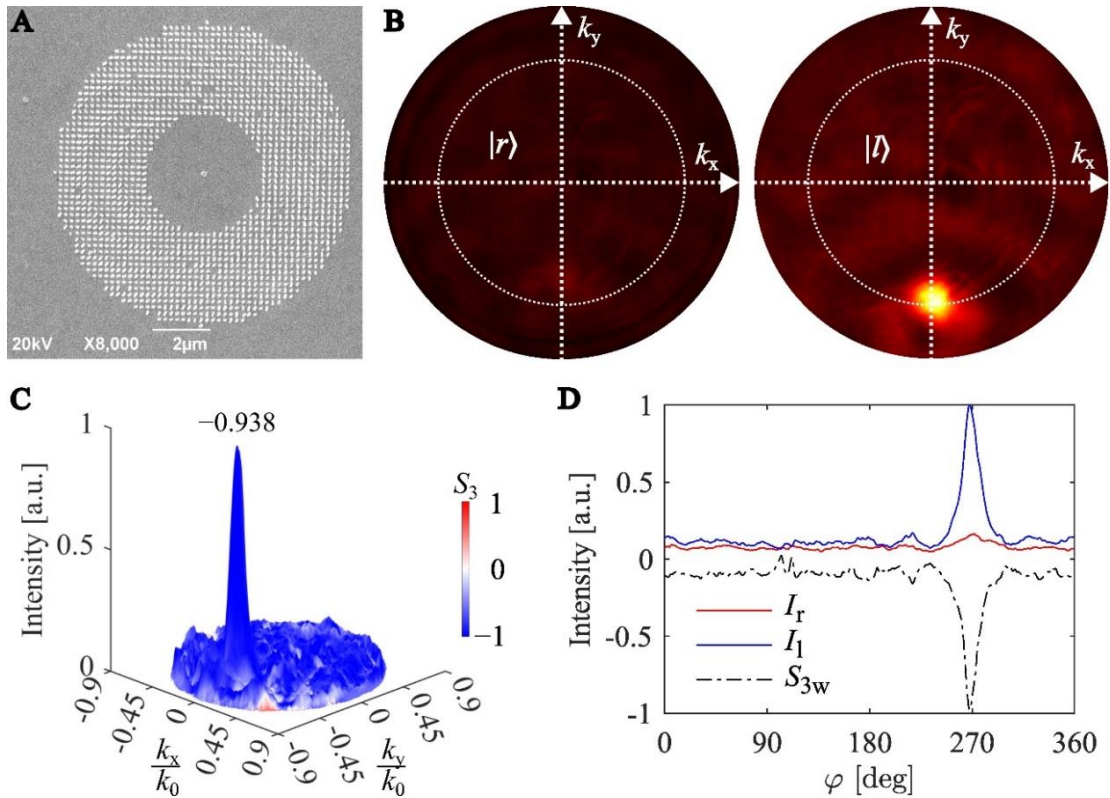

**Figure 3:** Experimental demonstration of well-collimated, single-channel emission with $|l\rangle$ state. (**A**) SEM image of the fabricated metasurface with a diameter of 10 μm and an inner radius of ~1.9 μm. (**B**) Polarization-resolved far-field intensity distributions of orthogonally polarized emission. Dashed circles indicate the emission polar angle of $\theta = 30°$. (**C**) 3D representation of the superposed far-field intensity (height) and Stokes parameters $S_3$ (color). (**D**) Cross-sectional distributions of the normalized intensities $I_l$ and $I_r$ and weighted Stokes parameters $S_{3w}$ on a circular path corresponding to the polar angle of $\theta = 30°$.

**Experimental demonstration of well-collimated, multichannel photon emission**

Phase matching two CP components with different deflection angles results in a metasurface (Fig. 4a) that effectively splits the photon emission into distinct $|r\rangle$ and $|l\rangle$ spots at $(\theta_r, \varphi_r) = (30°, 180°)$ and $(\theta_l, \varphi_l) = (30°, 90°)$ in the far field with equal intensities, as shown in Fig. 4b. The fabricated metasurface demonstrates $\overline{S_3}$ values of −0.942 for the $|l\rangle$ channel and 0.906 for the $|r\rangle$ channel, indicating high degrees of circular polarization for both CP channels (Fig. 4c). Another innovative design possibility in our approach is to create multichannel LP emission with qual intensities but orthogonal polarization states and arbitrary directionality by matching the phase distributions of multiple outgoing LP photons simultaneously. For instance, we implemented a metasurface that emits two spatially separated beams with $|a\rangle$ and $|b\rangle$ [$|b\rangle = \frac{1}{\sqrt{2}}(-|x\rangle + |y\rangle)$] states in the far field, as depicted in Fig. 4d-f. This approach successfully generates two orthogonal linear polarization states with high purity, achieving the averaged Stokes parameter $\overline{S_2}$ of 0.946 and −0.936 for the $|a\rangle$ and $|b\rangle$ states, respectively (Fig. 4f). Despite the impressive performance, this multichannel approach shows a decreased contrast between the two orthogonal polarization states when compared to single-channel designs. However, simulation results indicate that this reduction in contrast is not due to the metasurface design itself but is likely attributable to fabrication deviations (Supplementary Fig. S12). The calculated EQE in the multichannel emission is approximately 0.61, highlighting the efficiency of this design in generating desired photon sources.

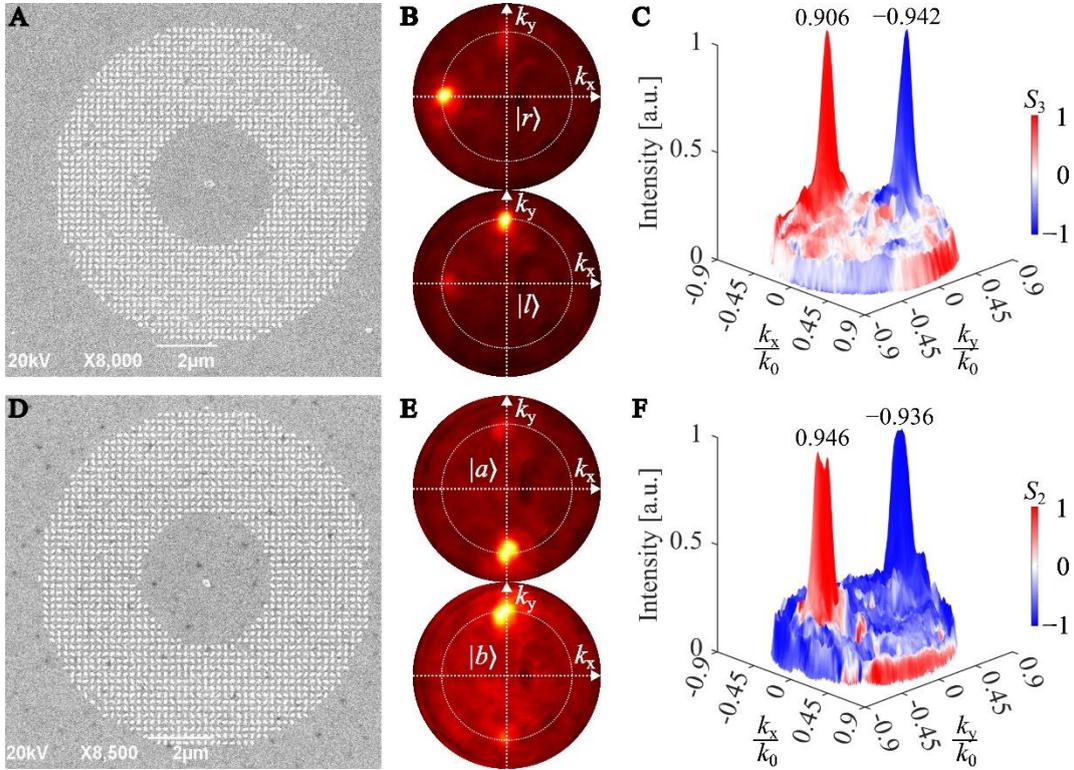

**Figure 4:** Experimental demonstration of well-collimated, multichannel photon emission with equal intensities. (**A, D**) SEM images of the fabricated metasurfaces that emit two spatially separated beams with (**A**)

$|r\rangle$ and $|l\rangle$ as well as (**D**) $|a\rangle$ and $|b\rangle$ states, respectively. The metasurfaces have diameters of 10 µm and inner radii of ~1.9 µm. (**B, E**) Polarization-resolved far-field intensity distributions of orthogonally polarized emissions. Dashed circles indicate the emission polar angle of $\theta = 30°$. (**C, F**) 3D representations of the superposed far-field intensities (height) and Stokes parameters (**C**) $S_3$ and (**F**) $S_2$ (color), respectively.

To fully utilize the DoFs offered by this design approach, we demonstrated a metasurface capable of generating $|a\rangle$ and $|l\rangle$ photons with distinct directions and adjustable relative intensities, as illustrated in Fig. 5. Initially, when the weights for each emission channel are set to 1, a lower intensity for the $|l\rangle$ channel was observed compared to the $|a\rangle$ channel. This discrepancy can be attributed to the fact that in the $|l\rangle$ channel, only one CP component is phase-matched and emitted out, whereas both CP components contribute to the $|a\rangle$ polarization state channel, resulting in an intensity ratio of 2:1 between the $|a\rangle$ and $|l\rangle$ channels. However, by adjusting the weights, it is possible to reverse this intensity distribution and achieve the desired balance between the two channels. The experimental results indicate consistent intensity distributions with the simulations (Supplementary Fig. S13). For polarization purity, the $|l\rangle$ channels exhibit large averaged $\overline{S_3}$ values of −0.918 and −0.928 in the experiment, consistent with simulated values of −0.990 and −0.990, respectively. In contrast, the $|a\rangle$ channels indicate decreased $\overline{S_2}$ parameters of 0.789 and 0.792 in the experiment, and 0.961 and 0.929 in the simulation. These results highlight the high purity of the circular polarization state in the $|l\rangle$ channels and the moderately lower purity in the $|a\rangle$ channels. One factor that causes the decreased purity of the $|a\rangle$ channel compared to the $|l\rangle$ channel is that the outcoupled LCP light contributes to both polarization states, whereas the RCP component only contributes to the $|a\rangle$ channel. This may lead to unequal intensities of RCP and LCP components in the $|a\rangle$ channel and results in decreased polarization purity.

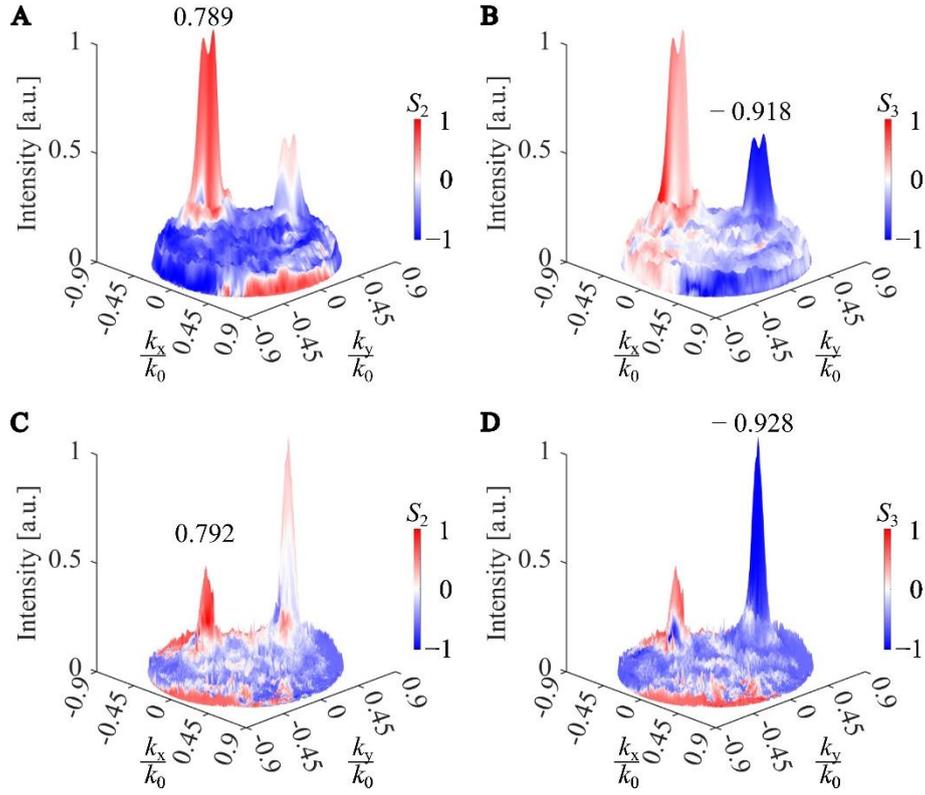

**Figure 5:** Experimental demonstration of well-collimated, multichannel photon emission with versatile polarization states and adjustable intensities. (**A, B**) 3D representation of the superposed far-field intensity (height) and Stokes parameters (**A**) $S_2$ and (**B**) $S_3$ (color) for $|a\rangle$ and $|l\rangle$ multichannel emission with an intensity ratio of 2:1. (**C, D**) 3D representation of the superposed far-field intensity (height) and Stokes parameters (**C**) $S_2$ and (**D**) $S_3$ (color) for $|a\rangle$ and $|l\rangle$ multichannel emission with an intensity ratio of 1:2.

## Conclusion

In this work, we demonstrated a universal approach to designing on-chip QE-coupled meta-optics that allow for simultaneous control over the polarization and directionality of photon emission, effectively overcoming the limitations of previous metaphotonic designs for versatile single- and multichannel quantum light sources. We developed a comprehensive theoretical framework and detailed the mechanisms by which the phase modulation of HWP meta-atoms enables the direct transformation of QE-excited SPPs into spatially propagating photon beams with full DoFs. Our experimental results showcased the capability of QE-coupled meta-optics in generating highly directional and off-normal single-channel photon emissions with pure linear (e.g., $|x\rangle$, $|y\rangle$, and $|a\rangle$) and circular ($|l\rangle$) polarization states. Moreover, we extended our design to achieve multichannel emissions, including angularly separated orthogonal polarization states ($|l\rangle$ and $|r\rangle$ as well as ($|a\rangle$ and $|b\rangle$)) with equal intensities. Additionally, we demonstrated the ability to generate versatile polarization states ($|a\rangle$ and $|l\rangle$) with tailored intensities by adjusting the weights of each emission channel. Beyond these capabilities, our platform demonstrated the potential to create complex emission profiles, such as vector beams and vortex beams, by incorporating more sophisticated phase profiles. This versatility is essential for advanced photonic applications, where precise control over emission properties is crucial. Our findings underscore the significant potential of phase-matched metasurfaces in producing high-fidelity, multichannel polarized photon emissions with simple adjustments to the design parameters, paving the way for future advancements in both classical and quantum information technologies.

## Data availability

The simulation and experiment data that support the findings of this study are available from the corresponding author upon request.

## Acknowledgments

This work was supported by the Villum Fonden (Grant No. 37372 and Award in Technical and Natural Sciences 2019) and Danmarks Frie Forskningsfond (1134-00010B). Y.K. acknowledges the support from European Union's Horizon Europe research and innovation programme under the Marie Skłodowska-Curie Action (Grant agreement No. 101064471).

quantum metasurfaces. *Adv. Mater.* **10**, 2212244 (2023).